# Design and development of a freeform active mirror for an astronomy application

Zalpha Challita
Tibor Agócs
Emmanuel Hugot
Attila Jaskó
Gabby Kroes
William Taylor
Chris Miller
Hermine Schnetler
Lars Venema
Laszlo Mosoni
David Le Mignant
Marc Ferrari
Jean-Gabriel Cuby

# Design and development of a freeform active mirror for an astronomy application

Zalpha Challita,[a,*] Tibor Agócs,[b] Emmanuel Hugot,[a] Attila Jaskó,[c] Gabby Kroes,[b] William Taylor,[d] Chris Miller,[d] Hermine Schnetler,[d] Lars Venema,[e] Laszlo Mosoni,[c] David Le Mignant,[a] Marc Ferrari,[a] and Jean-Gabriel Cuby[a]
[a]Aix Marseille Université, CNRS, LAM (Laboratoire d'Astrophysique de Marseille) UMR 7326, 13388, Marseille, France
[b]NOVA Optical Infrared Instrumentation Group at ASTRON, P.O. Box 2, 7990 AA Dwingeloo, The Netherlands
[c]Konkoly Thege Miklós Astronomical Institute (MTA Research Centre for Astronomy and Earth Sciences), 1525, PO Box 67, Budapest, Hungary
[d]Astronomy Technology Centre (UK-ATC), Blackford Hill, Edinburgh, EH9 3HJ, Scotland
[e]ASTRON, P.O. Box 2, 7990 AA Dwingeloo, The Netherlands

**Abstract.** The advent of extremely large telescopes will bring unprecedented light-collecting power and spatial resolution, but it will also lead to a significant increase in the size and complexity of focal-plane instruments. The use of freeform mirrors could drastically reduce the number of components in optical systems. Currently, manufacturing issues limit the common use of freeform mirrors at short wavelengths. This article outlines the use of freeform mirrors in astronomical instruments with a description of two efficient freeform optical systems. A new manufacturing method is presented which seeks to overcome the manufacturing issues through hydroforming of thin polished substrates. A specific design of an active array is detailed, which will compensate for residual manufacturing errors, thermoelastic deformation, and gravity-induced errors during observations. The combined hydroformed mirror and the active array comprise the Freeform Active Mirror Experiment, which will produce an accurate, compact, and stable freeform optics dedicated to visible and near-infrared observations. © 2014 Society of Photo-Optical Instrumentation Engineers (SPIE) [DOI: 10.1117/1.OE.53.3.031311]

Keywords: freeform mirrors; active optics; optical design and manufacturing; deformable mirrors; actuators; system architecture.

## 1 Freeform and Active Optics

Freeform optics has the potential to drastically reduce the complexity of optical instruments. Several definitions of "freeform optics" can be found in the literature with the most common being that the shape of a freeform optics must be nonaxisymmetric.[1] It is a rather loose definition; mathematically, since an off-axis conical section could be described as a nonaxisymmetric optics, as was demonstrated by Lubliner and Nelson[2] as far back as 1980. Likewise, adaptive and active optics have been used for many years to create nonaxisymmetric optics through the addition of aberrations to the classical optical profiles. For freeform optics, the real difference and originality lie in the extreme deviation from the best-fit sphere, which can be as large as a millimeter. As will be discussed, such extreme shapes require different manufacturing techniques compared with those of producing classical optical components.

There are numerous technological benefits associated with freeform optics: a reduction of overall instrument mass and volume, an improvement in reliability and operational availability, and an increased throughput.[3,4] Many of these advantages would be especially evident in future extremely large telescopes, which look set to house some of the largest instruments yet built.[5–7] However, despite these advantages, the implementation of freeform optics is currently held back by the difficulty of manufacturing highly aspherical optical surfaces that are of sufficient quality to allow visible imaging.

In this article, we present the concept of FAME, the Freeform Active Mirrors Experiment, and report the current status and direction of the project. The goal of FAME is to produce a freeform mirror whose extreme aspherical shape is actively controlled by a dedicated array of actuators. The optical surface or "face-sheet" will be manufactured through hydroforming a prepolished thin mirror. However, this process, which exploits the plastic behavior of ductile materials, will not yield a perfect freeform mirror. While the high-order optical quality of the mirror will be secured by the prepolishing of the face-sheet, the low- and mid-order shape errors will be corrected by an active array using techniques not dissimilar to active optics.[8–10]

Through the use of active optics techniques, FAME aims to relax both the requirements on the face-sheet manufacturing and that of the optical alignment. The acceptable errors on the face-sheet can be defined in terms of spatial frequencies instead of amplitude. Considering an array of $N^2$ actuators, the system allows the correction of spatial frequencies up to $N/2$ cycles per pupil. The amplitude of required corrections is quantified by finite element analysis (FEA), which allows constraints to be placed on the required force and stroke of the actuators. Within a predefined limited range, the active array can adapt its shape to optimize or adjust the mirror's performance. In addition, the system will incorporate a metrology system that measures the performance of the optical train and calculates the required changes on the individual actuators.

The work presented in this article reports on the current status of various aspects of this ongoing project. It begins with a comparison of classical designs to those which include freeform optics. It then details the FEA of the hydroforming

---

method and presents the results of the prototyping of the freeform face-sheet and the active array with preliminary analysis of the manufacturing method's performance. Finally, a discussion on the required control/command strategies is presented.

## 2 Optimizing Active Freeform-Based Optical Designs

### 2.1 Method

Optical designs that make use of freeform mirrors are becoming more common. In illumination systems, freeform optics has been used for some time. However, for more complex problems, where it is necessary to form images, new design tools are required. One such tool is the simultaneous multiple surfaces modeler. Nodal-aberration theory[11] and other theories regarding nonaxially symmetric systems are also valuable and have facilitated different design methods. Several polynomial families have been developed to mathematically represent the freeform surface such as the Bernstein,[12] the Forbes,[13] or the $\varphi$-polynomials.[1] These polynomials provide more freedom to describe the surfaces compared with that of the standard Zernike polynomials.

Usually, the performance metric used in the optimization process of an optical system is based on better image quality, faster $F$-ratio, larger field-of-view (FoV), or a combination of these. A limited space envelope or accessibility requirements are strong drives to use more complex optical surfaces, although freeform optical components generally deliver a system with superior characteristics.

The improvement of a system can only be made efficiently when the coupling between the different degrees of freedom (DOFs) of a system and the effect on the performance metric is well understood. For freeform optics, the DOF (the configuration vector) increases and results in a more complex optimization algorithm.

The correlation between the configuration vector and the performance metric can be written as $m = f(x)$, where $m$ is the performance metric, $x$ is the configuration vector, and $f$ is a function that defines the relationship. We intend to find the best $x$ to minimize $m$. Typically, there is a maximum allowable value for $m$, corresponding to the requirements of the system. Usually the function $f(x)$ is nonlinear, and there are interdependencies between the configuration parameters, which result in an ill-defined optimization problem. We developed the method described in Fig. 1 to overcome these difficulties and to minimize them. The key steps in the method are as follows:

- Singular value decomposition (SVD)-based subset selection[14,15] through the SVD of the sensitivity matrix (the change in performance by small variations in the configuration vector). This reduced the set of optimization parameters, which is then used to optimize the design.
- We used Newton's method based on SVD and the standard optimization methods of Zemax® that are based on damped least squares and orthogonal descent to optimize the design and as such to minimize the performance metric.

As usual, the optical design process aims to find the global minimum of the solution space and tries to avoid the local minima or divergence. The optical design method presented above optimizes the solution using the most sensitive configuration vector subset. Essentially, the optimization method intends to smooth out the surface of the performance metric by changing—in each optimization cycle—the variables of the system based on their current influence on the performance metric.

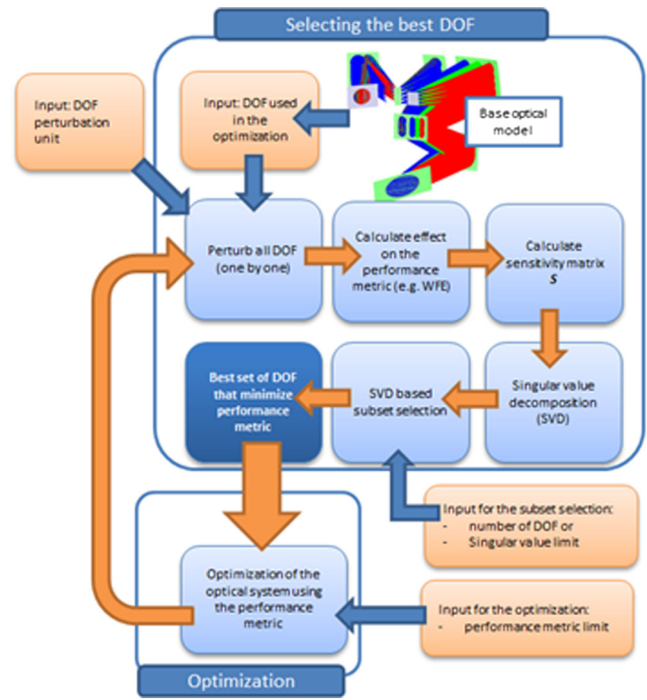

**Fig. 1** Optical design methodology used in the design of freeform optical components.

### 2.2 Comparing Freeform and Classical Optical Designs

In the following section, we present optical designs using extreme freeform surfaces. The main aim of the optimization process is to design an optical system that shows an improvement in some of the characteristics compared with the initial (classical) design. Also, more importantly, the design shall consist of only one extreme freeform surface, and the rest of the optical components have to be spherical or at least should be easy to manufacture and test. This configuration has a huge benefit in that the optical system itself can be used as the test-bed for the freeform surface, since the spherical surfaces can be manufactured and aligned with high accuracy.

In this section, we describe the design of two optical systems making use of freeform surfaces: a modified Three-Mirror Anastigmat (TMA) and a very wide-field and fast $F$-ratio two-mirror imaging system. We kept the same entrance pupil and the $F$-ratio during the reoptimization for ease of comparison while using the root mean-squared (RMS) spot radius as the performance metric. The aim of using freeform surfaces was to improve the optical performance and also to increase the FoV. The designs are shown in Fig. 2.

For the first freeform system, the starting point is a TMA that resembles the Cook design and has an 8 deg × 8 deg

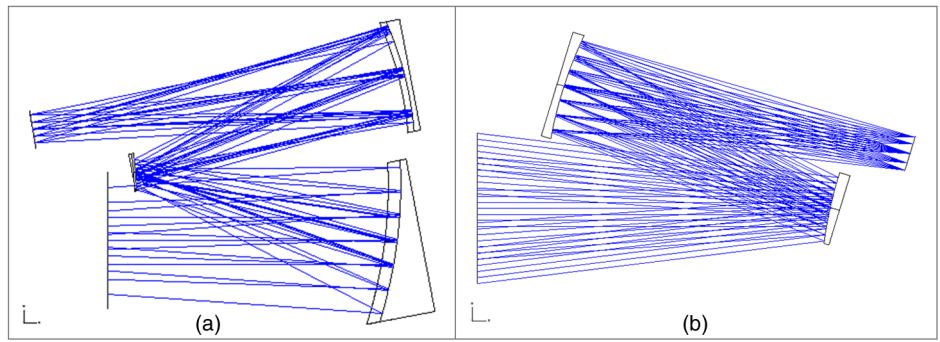

**Fig. 2** (a) Layout of a F4 Three-Mirror Anastigmat (TMA) design that resembles the Cook design with the freeform primary, spherical secondary, and tertiary. It has a field-of-view (FoV) of 8 deg ×12 deg (the initial design with three conical sections provides 8 deg ×8 deg FoV). (b) Layout of the freeform based two-mirror system (10 deg ×10 deg).

FoV. The primary mirror proved to be the best candidate for a more complex surface with higher DOF.

During the optimization process, it became apparent that the best performance can be achieved with a design that has the stop located close to the secondary mirror. It was possible to increase the FoV to 8 deg × 12 deg with an image quality that is better than the starting-point design.

The second optical design is a very wide-field two-mirror camera which would be suitable for large surveys with minimal observation time due to the high transmission. This optical design better suits a possible prototype, since it is a relatively simple system with only two components and is therefore easier to align. An off-axis Schwarzschild optical system was the starting point of the optimization with an entrance pupil of 50 mm, a FoV of 10 deg ×10 deg and an $F$-ratio of 4. The end-point design had a convex primary freeform mirror with a diameter of 100 mm. It is an off-axis asphere with the addition of Zernike polynomials (which corresponds to a freeform shape or $\varphi$-polynomials definition), and the second mirror is an ellipsoid with minor asphericity. Compared with the initial (classical) design, image quality improved significantly: the RMS spot radius of the design is improved by almost 40%.

In Fig. 3, both the sag and the deviation from the best fit sphere (BFS) for the primary mirrors are shown. For the TMA, in the case of the end-point design, it can be seen that the primary mirror has a 1.8-mm departure from BFS, so it is an extremely difficult surface to manufacture with conventional methods.

## 3 Hydroforming Freeform Shapes

Manufacturing freeform mirrors that satisfy the requirements for visible and infrared observations are a real challenge. As described by Fuerschbach et al.[1] in 2011, these components are more suited for thermal infrared (LWIR band) observations. Based on the preliminary specifications of the two optical designs described in the previous section, the FAME manufacturing method aims to deliver the required optical performance, while also keeping the manufacturing costs low. Nevertheless, the development of a brand new method raises several issues that need to be addressed in detail.

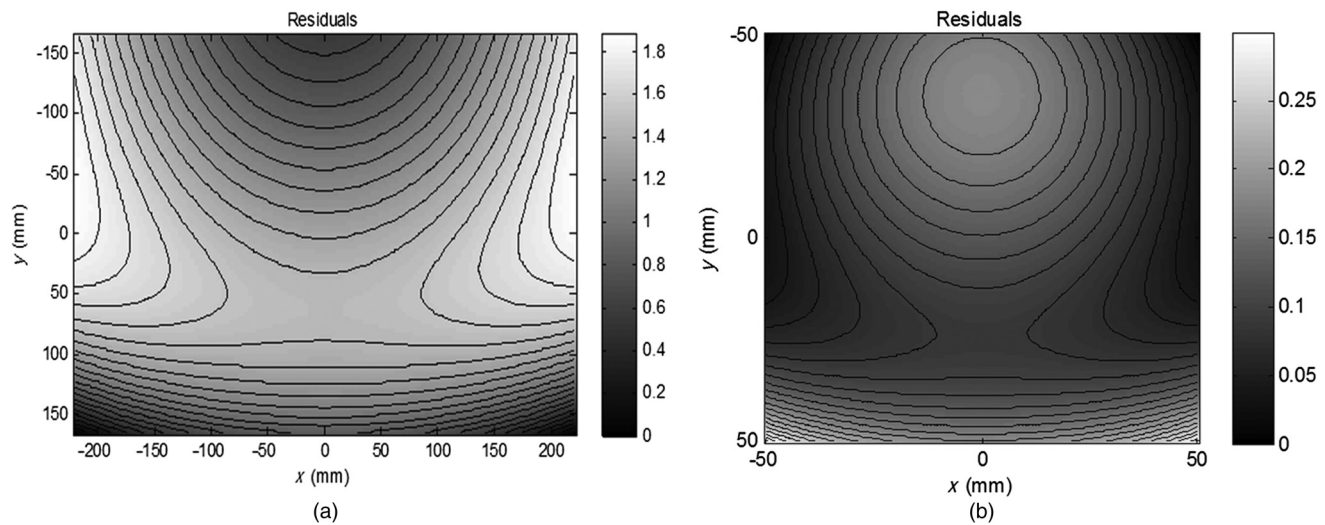

**Fig. 3** Deviation from best fit sphere (BFS). (a) Primary mirror of the F4 TMA design. (b) Primary mirror of the two-mirror design. In the first case, the mirror is an extreme freeform with 1.8-mm deviation from the BFS. (Units on both figures are in mm.)

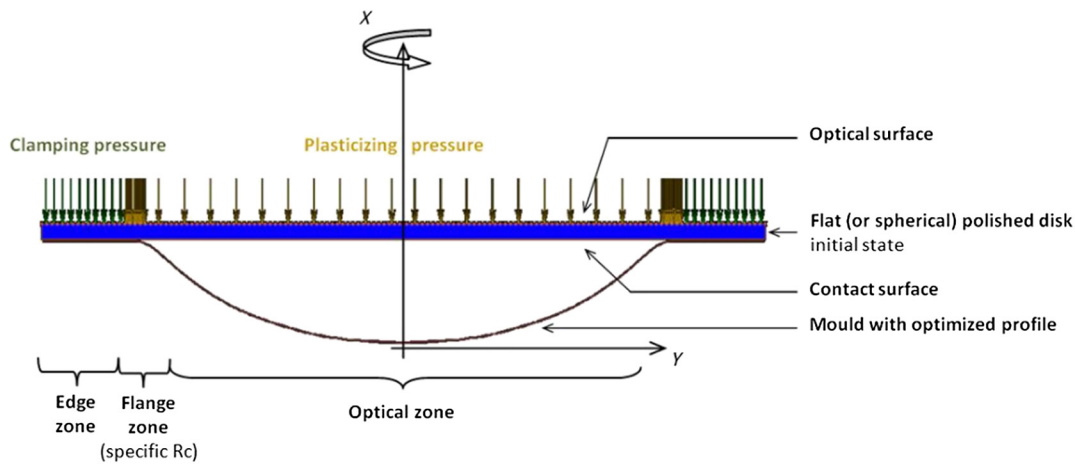

**Fig. 4** Substrate and mold functional aspects.

### 3.1 Plasticizing of Metallic Materials

The hydroforming method obtains freeform shapes through deformation of flat or spherical thin metallic mirrors. Such substrates have principally been used for the aluminum dishes required for radio-astronomy applications.[16] The hydroforming method is illustrated in Fig. 4. The prepolished substrate is held at its edges on a mold with a specific profile, and then a large amount of pressure is applied by a fluid until the substrate is formed into the shape of the mold. The resulting substrate keeps a permanent deformation after the removal of the pressure, as the material enters the plastic domain of ductile materials. This technique avoids any hard contact between the optical surface and the mold, meaning that the small local errors of the mold do not degrade the polished optical surface. This process only imparts the low-spatial frequency deformations, whereas any high-spatial frequency errors are not imprinted on the mirror's surface—a key advantage of hydroforming. The final expected residual errors are only form errors that could be compensated for by an active array with a reasonable number of actuators.

Several metallic materials can be selected based on crucial parameters such as polishing ability, stability over time, elastic limit, and plastic domain. Based on our experience in the field of metallic mirrors, we selected stainless steel (AISI420), aluminum (Al 6061-T6), and titanium (TA6V, T40) as candidates for prototyping and while developing the manufacturing process. The development method was implemented in three steps: finite element predictions, prototyping, and comparison.

### 3.2 Modeling Methods and Study Cases

The FEA provides good approximations of the plastic behavior of the substrate, but the predictions are limited by unknowns such as: the history of the material, residual stresses, the initial and the final geometries of the substrate, work hardening, and anisotropic microstructure. The FEA results highlighted two possible issues: springback effect and the degradation of the surface. Having the ability to compare the predicted performance (from the model) with real tests is important to allow us to fine tune the model.

The polished substrate is defined by three functional zones, as is shown in Fig. 4: the edge zone to hold the mirror in place, the flange zone where plasticization will occur, and the optical zone. A homogeneous controlled pressure is applied on the optical zone, whereas a clamping pressure is applied on the edge zone. These conditions are optimized using FEA.

As described in Fig. 5, two different approaches can be chosen to extract the working parameters of the hydroforming process from the FEA. We define $z(r)$ as the sag, $r$ as the mirror radial coordinate, $R_C$ as the radius of curvature, and BFS as the best fit sphere. The first method uses a simple spherical mold with which both spherical and aspherical mirrors can be made. The final shape depends on the material parameters, the mold aperture, the substrate geometry, and the pressures applied.

The second method uses a mold whose shape is optimized, while the rest of the parameters are fixed. The initial shape of this second mold corresponds to the final freeform shape and is iteratively modified to take into account the springback effect.[17]

The FEA is performed in order to start a parametric analysis and to extract two relevant cases for the study using method one. The face-sheets are modeled with quadrangle elements that allow a fine meshing of the mirrors section.[18] The analysis is truly nonlinear including large displacements, elastic-plastic behavior, and contact analysis. The material used for this parametric analysis is stainless steel (X30Cr13 or AISI420b) that can be polished. About 15,000 elements are used, giving a sampling of 1000 points on the optical surface for the optical quality extraction.

Figure 6 summarizes the results obtained in terms of RMS and peak-to-valley (PV) deviations from the BFS. Moderate asphericity (about 300 $\mu$m) is evident on the 2-mm mirror when a spherical $F/0.5$ mold is used. The inverse case is highlighted with an $F/2$ mold and a 1-mm thick substrate: the resulting plasticized mirror remains spherical but presents a larger aperture (see also pictures in Fig. 7). At the end of the process, we expect to obtain a spherical mirror with an error of less than 2-$\mu$m RMS. Table 1 describes the working parameters and the optical quality obtained for these two particular cases.

### 3.3 Substrates Realization and Hydroforming Engine

The manufacturing process can be described in four main steps: machining of substrates, polishing, hydroforming,

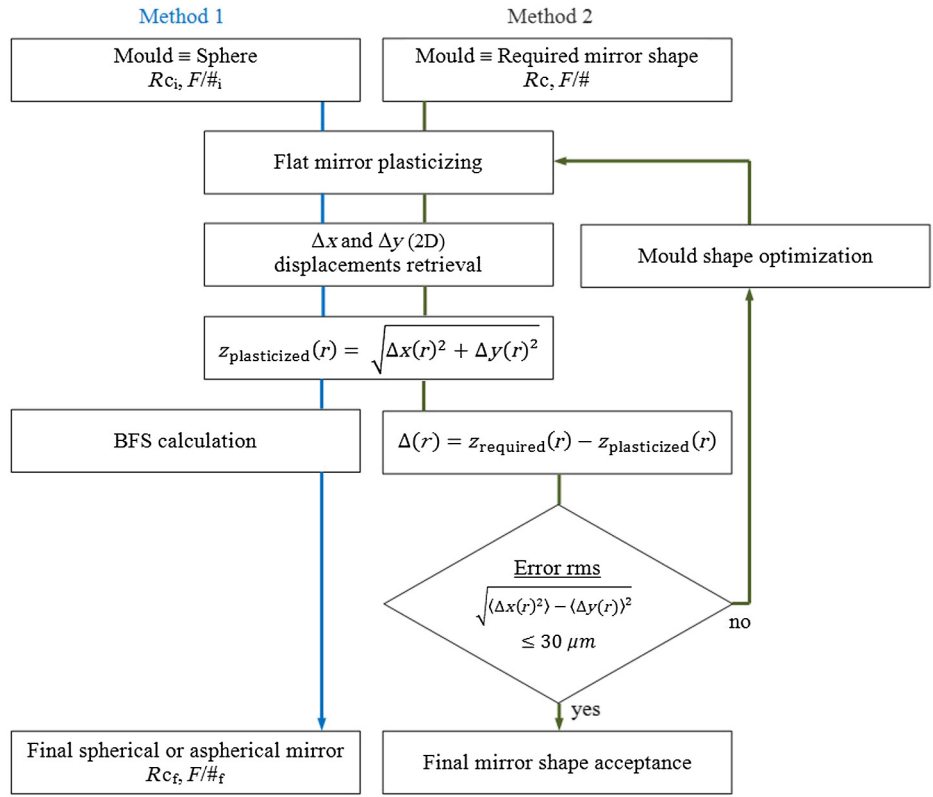

**Fig. 5** Two different optimization methods used to extract the working parameters.

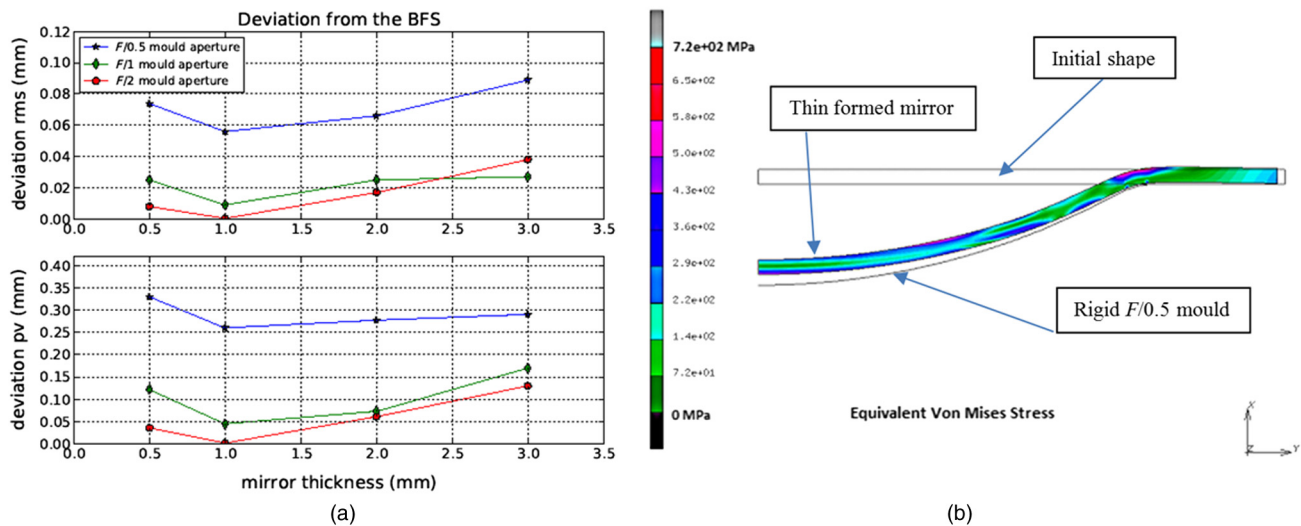

**Fig. 6** (a) For different mold apertures, evolution of the root mean-squared (RMS) and peak-to-valley (PV) deviations from the BFS, considering mirror thicknesses between 0.5 and 3 mm (100-mm optical diameter AISI420b mirror). (b) Redistribution of equivalent Von Mises stresses after the formation of a 2-mm substrate on $F/0.5$ spherical mold aperture, illustrating the springback effect.

and finishing. Several issues appear at each step of the process such as the anisotropy of the substrates, the introduction of surface stress during the machining phase and their relaxation during the polishing phase, the bending of the facesheets on the polishing supports, the bending on the test supports, and the degradation of the roughness due to plastic deformation of the surface. All these issues were overcome during the prototyping phase.

### 3.3.1 Machining and polishing of substrates

The method chosen to manufacture the initial substrates results from a compromise between: minimizing material anisotropy, minimizing stress induced during machining, reaching of high-mechanical planarity, and parallelism on thin and large substrate diameters. These aspects are important to improve the homogeneity of the material microstructure and to reduce the noncontrolled deformations of the

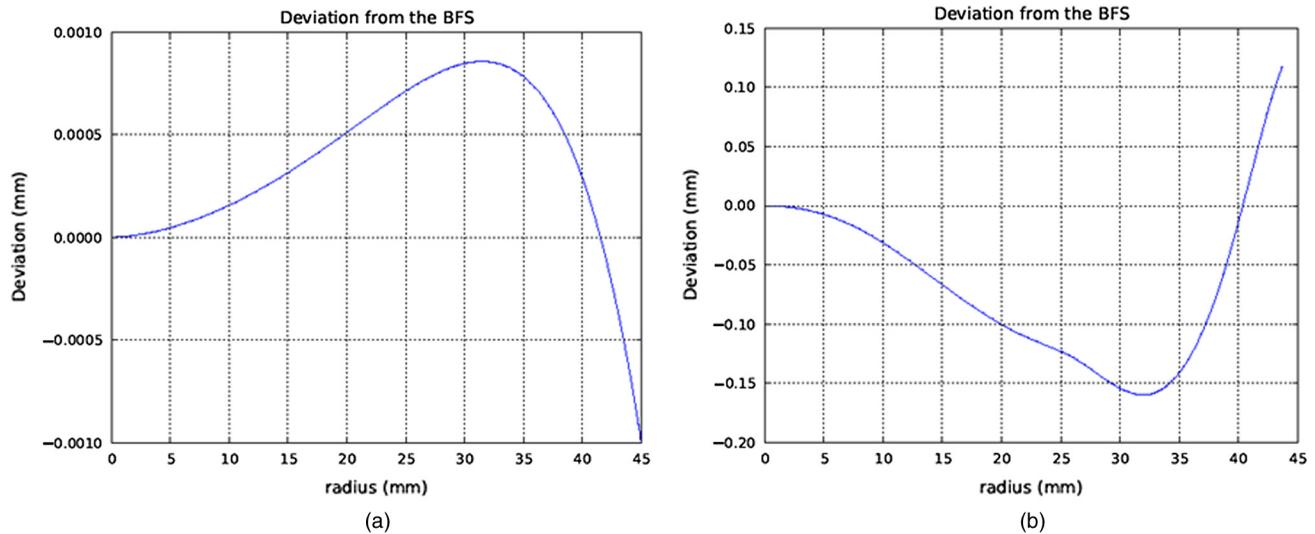

**Fig. 7** (a) Starting with a 1-mm thickness substrate on a spherical mold opened at $F/2$, the plasticized mirror remains spherical. (b) Starting with a 2-mm thickness substrate on a spherical mold opened at $F/0.5$, the plasticized mirror becomes aspherical with a departure from the BFS of 280 μm PV.

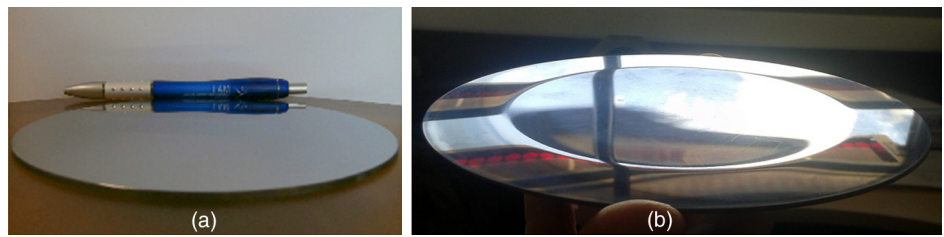

**Fig. 8** (a) 2-mm thickness mirror flat polished in AISI420b stainless steel. (b) The same mirror hydroformed.

**Table 1** Parameters of the two study cases. From a spherical mold, either spherical or aspherical mirror can be achieved.

| Study case | | Spherical | Aspherical |
|---|---|---|---|
| Working parameters | | | |
| Material | | X30Cr13 | X30Cr13 |
| Total disk diameter (mm) | | 140 | 140 |
| Optical diameter (mm) | | 100 | 100 |
| Disk thickness | | 1 mm, constant | 2 mm, constant |
| $F/$ mold | | $F/2$ | $F/0.5$ |
| Forming pressure (MPa) | | 15 | 45 |
| Clamping pressure (MPa) | | 10 | 10 |
| Optical quality obtained by modeling | | | |
| Error/departure from BFS | PV (μm) | 1.8 | 280 |
| | RMS (μm) | 0.38 | 66 |
| $F/D$ | | 10.7 | 0.54 |

piece due to stress relaxations. Obtaining thin and large diameter substrates on metallic materials of a high-mechanical quality was an issue tackled by a process using electrical discharge machining and double-sided surface grinding.

The substrates were machined from a round bar of AISI420b stainless steel. They present a total diameter of 140 mm and a thickness of approximately 1 and 2 mm with the following mechanical quality: flatness better than 30 μm RMS, a wedge lower than 20 μm RMS, and a 1.6 μm RMS roughness. A specific thermal treatment by annealing was also performed to relax residual stresses coming from machining and to ensure a stabilized mechanical behavior during polishing.

The polishing step is made easier as it is performed before the forming step, meaning it is reduced to a flat polishing. However, the polishing of high-aspect ratio metallic materials at optical level remains a difficult task. The polishing procedure developed at LAM, based on traditional full-sized tool polishing, allowed a flatness of 3 μm PV to be reached and a 5 nm RMS roughness. A flat polished mirror is illustrated on Fig. 8. The same mirror after hydroforming is also illustrated.

### 3.3.2 *Hydroforming engine*

Figure 9 presents the design of the hydroforming prototype. The flat polished mirror is integrated between the mold and

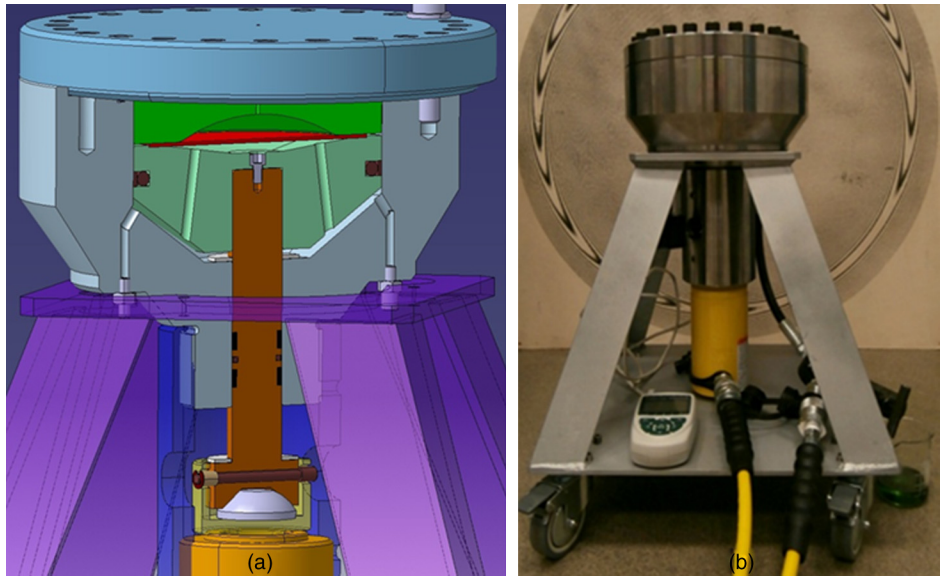

**Fig. 9** (a) Computer-aided design (CAD) of the hydroforming cavity. (b) Actual implementation of the hardware.

the clamping system inside a high-pressure cavity. Both the core and the hood of the cavity are made with hardened steel (as 30CND8, Re ∼950 MPa), thin enough to avoid elastic bending and to comply with the required high-pressure security coefficient. The clamping system is in contact with the edge of the mirror, and a jack allows pressure control. For safety reasons, the volume of fluid used for the forming pressure is minimized and shall not exceed 30 cl. At the end of the process, the applied pressures are removed, and the deformed mirror is extracted.

As presented previously, two structural effects have to be taken into account during plasticizing deformation: springback and roughness deterioration. The consequence of the springback effect is the introduction of residual form errors on the final freeform shape, despite the best approximations of the material parameters. These residual errors do not exceed 30 $\mu$m RMS and, due to the mirror's thinness, they can be corrected by *in situ* active compensation. This active compensation will also provide long-term shape stability, regardless of any relaxation of the material. In the next section, we present the work done thus far on the development of the active array.

## 4 Controlling Freeform Mirrors

### 4.1 Active Array for Freeform Mirrors Control

When it comes to designing an active array for a freeform mirror, the number of actuators, layout of array, and location of the actuators (grid) are important. The grid can have a standard equidistant repeating pattern, or an adaptable grid can be used to provide specific Eigen modes. When using an equidistant repeating grid, the number of actual grid points can be optimized with respect to the spatial order of the error. The grid form itself is independent from the aspherical shape, providing a common active array to create different surface shapes. This would allow the ideal optical train optimization and reconfiguration.

By providing a reconfigurable grid that follows the actual required slope/curvature of the surface, it will be possible to optimize the number of adjustment points, the shape of the influence functions, and the resulting Eigen modes. The performance of the active array is intrinsically related to the shape accuracy of the optical surface. Low-order modes are targeted, thus optimizing the DOF, reducing the cost, and simplifying the complexity of system control.

The control strategy for FAME is achieved by first correcting the low-order manufacturing errors, and then by correcting the thermoelastic effects due to environmental conditions. Both permanent and temporary deformations are required. A number of actuation topologies can be used: manual and/or motorized actuation or a combination of both.

### 4.2 Active Array Prototypes

After studying a number of design options,[19] like the piston or slope/curvature adjustment direction and the square/hexagon/triangular grid geometry, we came to the conclusion that for this application, an ideal active array has the following features: isostatic overall mirror mounting, internal slope/curvature adjustments, equidistant grid, adjustments between the nodes, and position-driven actuation.

A practical approach based on fundamental optomechanical engineering standards and rules was taken rather than trying to optimize the design in the theoretical domain. A series of "simple" prototypes formed the basis for the FEA validation, and the results are being used to define the final design. By comparing the measurements done on the prototypes with the FEA results obtained, the model can be optimized. The model can then be turned into a tool that can be used to predict the influence function for each actuation chain and to extract the Eigen modes of the complete freeform active mirror.

The goal is to determine the required actuations to accurately achieve a specific surface shape using the aforementioned tool. The first fundamental design solutions to perform the "internal" slope and curvature adjustments are shown in Fig. 10. The localized adjustments resulted in a smooth deformed surface.

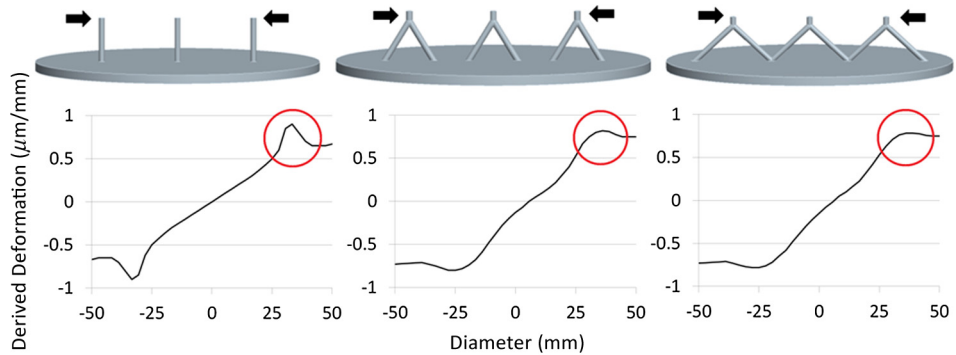

**Fig. 10** Pillar shape comparison between surface shapes and derivatives of the cross-section curves.

An FEA was performed on various pillar designs to provide more detailed information regarding the expected behavior and to help in the selection of the optimum shape. The pillar shape was changed in three steps from a single pillar to a pyramidal shape, as shown in Fig. 10, and the prototypes were depicted in Fig. 11. The surface shape and the derivatives of the cross-section curves were compared.

The first prototypes demonstrated a very good correlation between measured and predicted values (within a few microns). The PV deformations, at the cross-section through the input nodes, showed similar size and shape between the tests and FEA. The derived PV deformation curves showed that the actual slope for each point along the cross-section is very similar, and the results between tests also showed good correlation with the predicted FEA results. Linearity and hysteresis, two important effects, were tested with the pyramid-style prototype. The measurements showed a linear response with very small hysteresis.

In the next phase of the development, a two-dimensional adjustable array using a pyramid-like shape design will be designed and optimized. The array will be characterized by measuring the Eigen modes and optical performance.

### 4.3 Control Electronics and Evaluation

The last step required to produce a freeform active mirror is to integrate the active array with the control and drive electronics. The control and drive electronics are only required for the automated control loops. The proposed closed-loop control architecture is depicted in Fig. 12.

The design of the control and drive electronics is highly dependent on the functions required. Initially, the overall shape of the mirror, which might require larger corrections, will be performed manually. A second-order control loop will then be used to ensure shape stability of the mirror under all operational environmental conditions during deployment.

Most likely, temperature and own mass flexure will have the greatest influence on the stability of the mirror shape. The typical operational temperature for near-infrared astronomical instruments varies between 40 and 190 K. Thus, it is important to select actuators that can operate at those temperatures.

Although a wide temperature range has been specified, any one component will be typically operated at a very specific temperature with stability in the order of 2 K. In theory, the variations of the mirror shape due to temperature will be very small.

On the other hand, the variations due to flexure when an instrument has to track an astronomical object over the sky can result in relatively large deformations due to own weight flexures. To compensate for these gravity-induced errors, the system must be monitored with internal sensors such as resolvers, encoders, linear-variable differential transformers, or capacitive sensors for measuring displacements with nanometer accuracy.

Once the mirror has been characterized and the design of the active array has been finalized in terms of its requirements, the appropriate actuators and sensors will be selected. The design of the driving electronics will be dictated by the mechanism train and actuator type. Ideally, the actuators should dissipate as little power as possible to ensure that these mirrors do not act as hot spots within the optical train. The actuators must also be stable when the system is powered down.

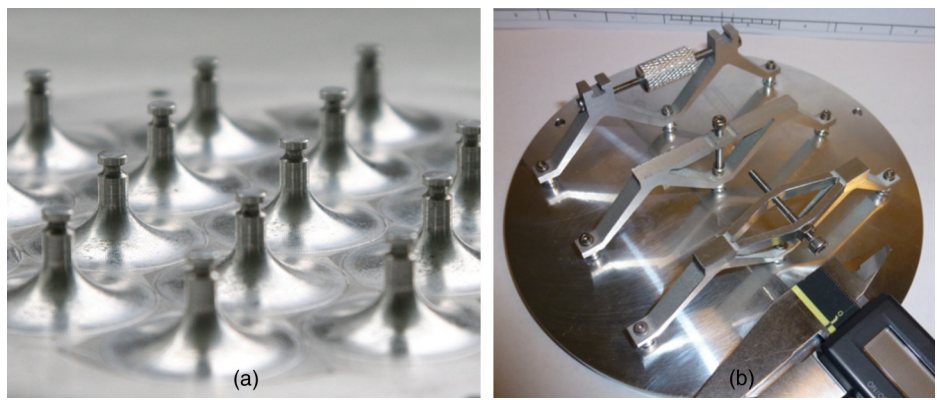

**Fig. 11** (a, b) Two active array prototypes.

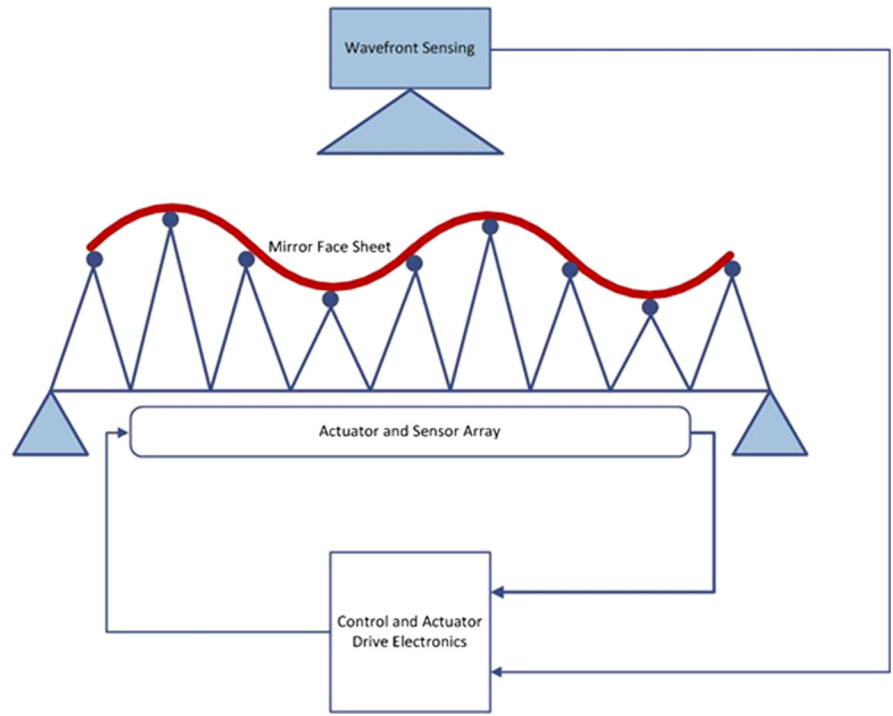

**Fig. 12** Freeform active mirror control and driving electronics.

## 5 Conclusion and Perspectives

Freeform mirror technology represents a breakthrough in terms of enabling large, lightweight, and compact telescope design. In this article, we have given an overview of the work done in three areas of active freeform mirror development. These study results will be used to drive the production of the first active freeform mirror prototype, thus demonstrating a novel technology that has the potential to decrease the size, weight, and number of optical surfaces required, while also increasing the FoV and image quality of the optical designs of the future astronomic instruments.

Through the use of freeform optical components, we demonstrated possible improvements for two optical design examples: both reflective and off-axis; each optimized to provide a wider FoV (several degrees) and better optical quality. We have presented a new hydroforming manufacturing process and also detailed the issues associated with this technique. The method developed utilizes the ability of thin metallic sheets to be permanently deformed by working the materials above their elastic limit through cold-forming techniques. Despite the lack of predictable parameters in the plastic domain, we present a method based on nonlinear FEA and a prototyping that allows us to rapidly converge to an aspherical shape where residuals are in the range of tens of microns.

The deviation in shape and form of the actual manufactured mirror from that of the "as-designed" shape will be corrected by introducing an active array, where the number and position of actuators could be adapted depending on the nature and size of the manufacturing errors. While the active array would primarily be employed to correct the manufacturing errors, it also has the potential to compensate for misalignment errors, thermoelastic deformations, and instabilities due to environmental variations. These corrections could be applied to the freeform mirror itself and could also compensate for errors in other components in an optical system, and thus dramatically increase the usefulness of systems by making use of freeform mirror that FAME develops.

*Acknowledgments*

This research and development project is partly funded by the European FP7-OPTICON WP5 "Freeform Active Mirrors Experiment" project and a grant from the region Provence-Alpes-Côte d'Azur, FEDER and Thales-SESO.

*References*

1. K. Fuerschbach, J. P. Rolland, and K. P. Thompson, "A new family of optical systems employing $\varphi$-polynomial surfaces," *Optics Express* **19**(22), 21919–21928 (2011).
2. J. Lubliner and J. E. Nelson, "Stressed mirror polishing. 1: a technique for producing non-axisymmetric mirrors," *Appl. Opt.* **19**(14), 2332–2340 (1980).
3. J.-G. Cuby et al., "On the performance of ELT instrumentation," *Proc. SPIE* **6269**, 62691U (2006).
4. O. Guyon et al., "Exoplanet imaging with a phase-induced amplitude apodization coronagraph. I. Principle," *Astrophys. J.* **622**(1), 744–758 (2005).
5. H. M. Martin et al., "Production of 8.4 m segments for the Giant Magellan Telescope," *Proc. SPIE* **8450**, 84502D (2012).
6. C. Baffes et al., "Primary mirror segmentation studies for the Thirty Meter Telescope," *Proc. SPIE* **7018**, 70180S (2008).
7. The E-ELT construction proposal, book_0046, credit ESO (2011).
8. G. R. Lemaitre, "Optical design and active optics methods in astronomy," *Opt. Rev.* **20**(2), 103–117 (2013).
9. C. Cunningham et al., "Smart Instrument technologies to meet extreme instrument stability requirements," *Proc. SPIE* **7018**, 70181U (2008).
10. R. N. Wilson, "Active optics and the new technology telescope (NTT): the key to improve optical quality at lower cost in large astronomical telescopes," *Contemp. Phys.* **32**(3), 157–172 (1991).
11. K. P. Thompson et al., "Using nodal aberration theory to understand the aberrations of multiple unobscured three mirror anastigmatic (TMA) telescopes," *Proc. SPIE* **7433**, 8 (2009).
12. S. Pascal et al., "New modelling of freeform surfaces for optical design of astronomical instruments," *Proc. SPIE* **8450**, 845053 (2012).
13. G. W. Forbes, "Robust, efficient computational methods for axially symmetric optical aspheres," *Opt. Express* **18**(19), 19700–19712 (2010).

14. H. N. Chapman and D. W. Sweeney, "A rigorous method for compensation selection and alignment of microlithographic optical systems," *Proc. SPIE* **3331**, 102–113 (1998).
15. G. H. Golub and C. F. Van Loan, *Matrix Computations*, 3rd ed., John Hopkins University Press, Baltimore (1996).
16. W. A. Imbriale et al., "The 6-Meter breadboard antenna for the deep space network large array," The Interplanetary Network Progress Report, Vol. 42–157, pp. 1–12, JPL, Pasadena, California (2004).
17. C. Bruni et al., "A study of techniques in the evaluation of backlash and residuals stress in hydroforming," *Int. J. Adv. Manuf. Technol.* **33**, 929–939 (2007).
18. Z. Challita et al., "Extremely aspheric mirrors: prototype development of an innovative manufacturing process based on active optics," *Proc. SPIE* **8450**, 845033 (2012).
19. G. Kroes et al., "A new generation active arrays for optical flexibility in astronomical instrumentation," *Proc. SPIE* **8450**, 845029 (2012).

**Zalpha Challita** is a PhD candidate at Laboratoire d'Astrophysique de Marseille, France. After a master's degree, oriented in space- and ground-based astronomical instrumentations, from the University of Paris XI-Orsay and the Paris Observatory, she pursues her activities in Research & Development dedicated to new instrumental concepts and new methods in optical fabrication.

**Tibor Agócs** received his MSc in engineering physics, specializing in optics, at the Budapest University of Technology and Economics, in 2003. Between 2003 and 2006, he worked as development engineer at Holografika, Budapest, Hungary. From 2006 to 2011, he was employed as an optical engineer at the Isaac Newton Group of Telescopes (ING) in La Palma, Spain. Since 2011, he has been employed as an optical designer at the NOVA Optical Infrared Instrumentation Group at ASTRON in Dwingeloo, The Netherlands. His main professional interest is optical design and analysis of optical systems in astronomical instrumentation.

**Emmanuel Hugot** is a researcher at Laboratoire d'Astrophysique de Marseille. Its R&D activities are dedicated to active optics and innovative instrumentation. Involved in several projects for ground-based observatories for years, he has now oriented his work on the high-angular resolution and high-contrast imaging for space missions, specifically through the development of innovative focal plane architectures based on active mirrors and freeform optics. He received a PhD in astronomical instrumentation from Aix-Marseille University, in 2007.

**Attila Jasko** is a mechanical designer at the Konkoly observatory of the Hungarian Academy of Sciences. He is working on astronomical instrumentation designs and their validation using finite-element method.

**Gabby Kroes** is lead mechanical engineer at the NOVA Optical InfraRed astronomical instrumentation group in The Netherlands. She works on the design and development of complex (cryogenic) instrumentation for ground-based (ESO VLT/VLTi) as well as space-based (ESA/NASA JWST) observatories. She specializes in the design of cryogenic optics mounts and cryogenic mechanisms. She received her BSc in precision engineering, in 1995.

**William Taylor** is currently based at the Royal Observatory Edinburgh, where he completed his PhD in 2012. His primary research involves furthering our understanding of the evolution of young massive stars and their host clusters. Instrumentation-wise, he has been heavily involved in the development of MAPS—a new mechanism whereby small mirrors are positioned using miniature robots. He has combined his scientific and technical background as part of instrument design studies for moons, eagle and echo.

**Hermine Schnetler** of SFTC-UK ATC is the lead system engineer for the CSP Consortium. With 20 years' experience as a systems engineer, she began in the defense industry on products such as inertial navigation systems for aircraft, helmet sighting systems, and helicopter-mounted sighting systems. She joined the STFC-UK ATC group 8 years ago and is the head of the systems engineering group. She tailored and successfully introduced a robust systems engineering process at the UK ATC and was also involved in a number of instrument studies. She is the chair for the International Council on Systems Engineering (INCOSE) Scottish Local Group, a member of the Institute for Engineering Technology and SPIE, and has developed a systems engineering course for astronomy projects, which she runs at the SPIE Astronomy Instrumentation Conferences.

**Lars Venema** is a senior scientist of optical instrumentation at ASTRON in Dwingeloo, The Netherlands, since 2006, having been the head of Optical Group of the R&D division of ASTRON during 5 years. After more than 12 years specialized in nuclear physics and geophysics, he is involved in a number of instruments projects dedicated to astronomy, in particular as a system engineer (METIS and EPICS-EPOL instruments phase-A studies).

**Laszlo Mosoni** studies the structure and composition of the protoplanetary disks of low-mass (i.e., sun-like) young stellar objects at high angular resolution with infrared stellar interferometers (e.g., ESO very large telescope interferometer). He also works in the field of interferometric image reconstruction.

**Marc Ferrari** is an astronomer at Laboratoire d'Astrophysique de Marseille. He received a PhD in astronomical instrumentation from Aix-Marseille University in 1994. After positions at ESA and ESO, he joined LAM in 2000. A specialist on active/adaptive optics and optical fabrication, he was the optics R&D group leader from 2004 to 2011. Since 2012, he has been deputy director at LAM.

Biographies of the other authors are not available.